\documentclass[
notitlepage,
amsmath,
floats,
floatfix,
twocolumn,
superscriptaddress,nofootinbib,showpacs,
usenames,dvipsnames,svgnames,x11names,
bookmarks=false,
pdfnewwindow=true,
final=true,
colorlinks=true,citecolor=RoyalBlue,filecolor=xlinkcolor,linkcolor=RoyalBlue,urlcolor=DarkGreen,
]{aastex62}

\usepackage{amsmath}
\usepackage{natbib}
\usepackage{marginnote}
\usepackage{comment}

\usepackage{microtype}

\usepackage[
        range-units=single,
        retain-explicit-plus=true,
        retain-unity-mantissa=false,
]{siunitx}

\DeclareSIUnit{\pc}{pc}
\DeclareSIUnit{\parsec}{pc}
\DeclareSIUnit{\year}{yr}
\DeclareSIUnit{\yr}{yr}
\DeclareSIUnit{\ppm}{ppm}
\DeclareSIUnit{\partspermillion}{ppm}
\DeclareSIUnit{\nothing}{\relax}
\DeclareSIUnit{\solarmass}{\text{\ensuremath{M_\odot}}}

\graphicspath{{figures/}}

\usepackage[capitalise]{cleveref}
\makeatletter
\usepackage{etoolbox}
\patchcmd\H@refstepcounter{\protected@edef}{\protected@xdef}{}{}
\makeatother

\newcommand{\mr}[1]{\mathrm{#1}}
\DeclareMathOperator{\re}{Re}

\newcommand{\SU}{\affiliation{Department of Physics, Syracuse University, Syracuse, New York 13244, USA}}
\newcommand{\MIT}{\affiliation{LIGO Laboratory, Massachusetts Institute of Technology, Cambridge, Massachusetts 02139, USA}}
\newcommand{\CIT}{\affiliation{LIGO Laboratory, California Institute of Technology, Pasadena, CA 91125, USA}}
\newcommand{\CSUF}{\affiliation{California State University Fullerton, Fullerton, CA 92831, USA}}
\newcommand{\PSUA}{\affiliation{Institute for Gravitation and the Cosmos, Department of Physics, Pennsylvania State University, PA 16803, USA}}
\newcommand{\PSUB}{\affiliation{Department of Astronomy and Astrophysics, Pennsylvania State University, PA 16803, USA}}
\newcommand{\CARDIFF}{\affiliation{School of Physics and Astronomy, Cardiff University, Cardiff CF24 3AA, United Kingdom}}
\newcommand{\CITA}{\affiliation{Canadian Institute for Theoretical Astrophysics, University of Toronto, Toronto, Ontario M5S 3H8, Canada}}

\submitjournal{ApJ}
\shorttitle{Science-Driven Tunable design of Cosmic Explorer Detectors
}
\shortauthors{Srivastava et al.}

\begin{document}

\title{Science-Driven Tunable Design of Cosmic Explorer Detectors}
\correspondingauthor{Varun Srivastava}
\email{vasrivas@syr.edu}

\author[0000-0002-4296-5463]{Varun Srivastava}\SU
\author[0000-0000-0000-0000]{Derek Davis}\CIT
\author[0000-0000-0000-0000]{Kevin Kuns}\MIT
\author[0000-0000-0000-0000]{Philippe Landry}\CITA
\author[0000-0000-0000-0000]{Stefan Ballmer}\SU
\author[0000-0000-0000-0000]{Matthew Evans}\MIT
\author[0000-0001-9018-666X]{Evan D. Hall}\MIT
\author[0000-0000-0000-0000]{Jocelyn Read}\CSUF
\author[0000-0000-0000-0000]{B.S. Sathyaprakash}\PSUA \PSUB \CARDIFF


\begin{abstract}

Ground-based gravitational-wave detectors like Cosmic Explorer can be tuned to improve their sensitivity at high or low frequencies by tuning the response of the signal extraction cavity. Enhanced sensitivity above $\SI{2}{\kHz}$ enables measurements of the post-merger gravitational-wave spectrum from binary neutron star mergers, which depends critically on the unknown equation of state of hot, ultra-dense matter. Improved sensitivity below $\SI{500}{Hz}$ favors precision tests of extreme gravity with black hole ringdown signals, and improves the detection prospects while facilitating an improved measurement of source properties for compact binary inspirals at cosmological distances. At intermediate frequencies, a more sensitive detector can better measure the tidal properties of neutron stars. We present and characterize the performance of tuned Cosmic Explorer configurations that are designed to optimize detections across different astrophysical source populations. These tuning options give Cosmic Explorer the flexibility to target a diverse set of science goals with the same detector infrastructure. We find that a $\SI{40}{\km}$ Cosmic Explorer detector outperforms a $\SI{20}{\km}$ in all key science goals other than access to post-merger physics. This suggests that Cosmic Explorer should include at least one $\SI{40}{\km}$ facility.

\end{abstract}



\section{Introduction}
\label{sec:Intro}

The next-generation of gravitational-wave detectors, Cosmic Explorer~\citep{Evans:2021gyd} and Einstein Telescope~\citep{ETBook, ETDesign2020}, are proposed to be operational by the mid-2030s. 
These detectors are expected to be 10 times more sensitive than current advanced LIGO \citep{{TheLIGOScientific:2014jea}} and Virgo \citep{TheVirgo:2014hva} observatories.
This allows the next generation of gravitational-wave facilities to observe compact binaries coalescences throughout the universe.
The observation of gravitational waves from diverse astrophysical sources opens avenues for novel scientific discovery and astrophysical understanding, which has been articulated in the Gravitational Wave International Committee third-generation (GWIC 3G) science Book~\citep{kalogera2021next}, the Einstein Telescope Science Case~\citep{MaggioreVanDenBroeck2020}, and the Cosmic Explorer Horizon Study~\citep{Evans:2021gyd}.
In particular, the Cosmic Explorer Horizon Study identifies three key science goals---mapping the cosmic history of merging black holes and neutron stars, exploring the nature of extreme matter through neutron star mergers, and testing fundamental physics and gravity in the strong-field regime. 
These science goals rely on the observation of gravitational waves from different astrophysical sources or astrophysical processes. 
The prospects of their observation depend on the sensitivity of Cosmic Explorer facilities in the relevant frequency band.

Signals at the low end of the frequency spectrum, below \SI{500}{Hz}, include black hole ringdowns, continuous gravitational waves from rotating neutron stars, and compact binary inspirals at cosmological distances. 
Because general relativity makes a precise prediction for the quasinormal modes of the remnant black hole ringdown formed after compact binary mergers, it allows for a critical test of Einstein's theory~\citep{PhysRevLett.116.221101,PhysRevD.100.104036,PhysRevD.103.122002,berti2018extreme2}. 
Continuous gravitational waves from isolated or accreting binary neutron stars carry information about crustal, thermal or magnetic deformations or internal mode excitations~\citep{Broeck_2005, Isi_CW_2017, Sieniawska_cw_2019}.
Observations of a large population of compact binaries at high redshift with precise source information is useful to constrain cosmological parameters, and trace the evolution of the compact binary populations, understand their formation channels and their progenitors across cosmic time~\cite{Abbott_2019_pop,Abbott_2021_pop}.

At intermediate frequencies from 500 to \SI{1500}{Hz}, tidal effects from neutron star mergers imprint on the gravitational waveform~\citep{Hinderer_2016,Dietrich:2017aum,Dietrich_2019}.
They reveal the internal structure of neutron stars, which tells us about the properties of zero-temperature supranuclear matter, namely its equation of state. More precise gravitational-wave measurements of neutron-star tidal deformability can advance our understanding of dense matter, especially in conjunction with electromagnetic observations of neutron stars~\citep{GW170817_Joint,Radice_2018,radice2019multimessenger}.

The post-merger gravitational waves from the oscillating remnants of binary neutron star coalescences lie at the high-frequency end of the spectrum, above \SI{2}{\kHz}. The post-merger oscillations depend sensitively on the structure and evolution of the hot, hypermassive neutron star remnant, which attains the highest matter densities in the Universe. Given that these signals are likely not detectable with Advanced LIGO and Virgo---even with so-called A+ technology or the proposed Voyager technology~\citep{LIGO:2020xsf}. However, Einstein Telescope, Cosmic Explorer and NEMO \citep{ETBook, ETDesign2020, Evans:2021gyd, Ackley:2020} will shed light on unexplored regions of the phase diagram of quantum chromodynamics by delivering reliable post-merger observations.

The large scale of the proposed third-generation gravitational-wave observatories endows them with broadband sensitivity from a few \si{Hz} to several \si{\kHz}.
This will not only enable them to capture the vast astrophysical population of known compact binary sources, but also opens the exciting possibility of unraveling new gravitational-wave sources, such as supernovae, isolated pulsars, or exotic compact objects.
Here we introduce a design for Cosmic Explorer that allows for tuning its sensitivity between observing runs to maximize its scientific output.
We present Cosmic Explorer tunings that optimize sensitivity to low-, intermediate- and high-frequency sources. 
The tuned configurations provide enhanced sensitivity in a frequency band that is optimized for detecting the corresponding astrophysical sources. 
This is particularly beneficial for future gravitational-wave detectors where increasing the circulating power will be challenging and may be technologically infeasible.

We discuss the tunable design of Cosmic Explorer in \cref{sec:tuning}. \cref{subsec:networks} provides a summary of the different gravitational-wave detector networks considered, and a summary of the tuned Cosmic Explorer configurations.
The configurations with an improved sensitivity at high frequencies are discussed in \cref{sec:high_freq_tuning}.
The section first summarizes the current understanding of the nature of post-merger signals in \cref{sec:pmr_signal}, and is followed by the post-merger tuning in \cref{subsec:pmr_tune}. 
Tuning focused to improve the measurement of the tidal deformability of binary neutron stars, and the corresponding improvement is summarized in section~\cref{subsec:tidal_tune}.
The low frequency tuning is discussed in \cref{sec:low_freq_tuning}, and the relative improvement in the detection prospects of sources at high redshift is summarized in \cref{subsec:snr_imp,subsec:tides_highz}. 
The impact on the observation of the ringdown of black hole remnants and the continuous wave sources is discussed in \cref{subsec:grav_tune} and \cref{subsec:cw}, respectively.
Limitations to these high and low frequency tuned configurations is discussed in \cref{sec:technology_drivers}.
\cref{sec:discussion} summarizes the key results of our paper.

\section{Tunable Design of Cosmic Explorer}
\label{sec:tuning}
The reference design and the technological advances required to achieve Cosmic Explorer's unprecedented sensitivity are discussed in the Cosmic Explorer Horizon Study~\citep{Evans:2021gyd} and in~\citet{Hall:2020dps}.
Unlike second-generation gravitational wave detectors~\citep{TheLIGOScientific:2014jea, TheVirgo:2014hva, Akutsu:2018axf}, Cosmic Explorer's design makes it feasible to optimize its sensitivity for a specific science goal with only minor modifications to the detector between observing runs.

\begin{figure}
  \centering
  \includegraphics[width=\columnwidth]{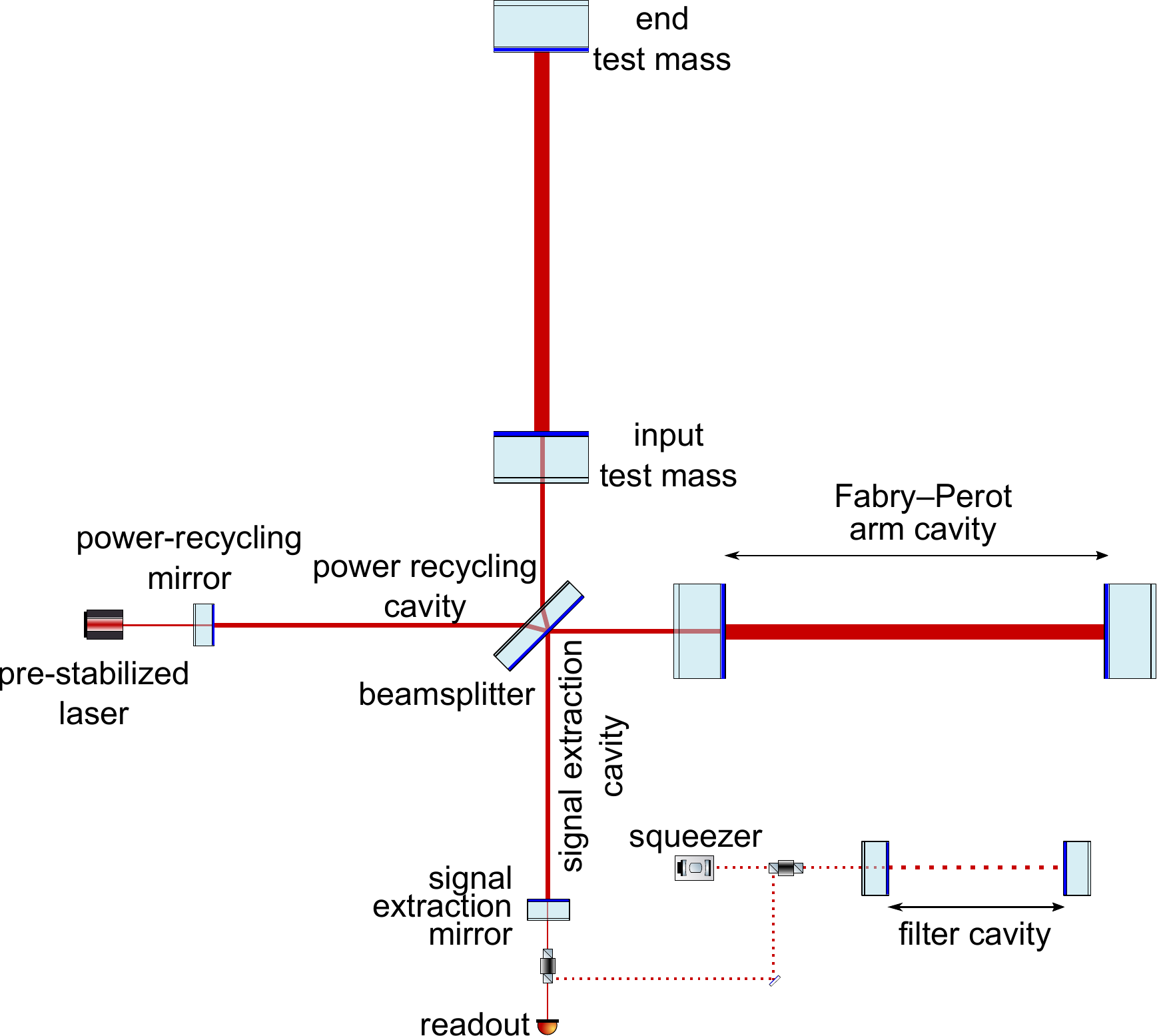}
  \caption{Simplified optical layout of the Cosmic Explorer interferometer. The signal extraction mirror forms a coupled cavity with the arms and its properties determine the shape of the detector's response. The design choices described here consist of 1)~choosing the length of this cavity when the facility is built, and then 2)~tuning the detector response by changing the signal extraction mirror between observing runs. The squeezer and filter cavity improve the quantum noise in the detector, but their design imposes no constraints on the tuning even though minor modifications to the filter cavity would be required when switching between tunings.}
  \label{fig:ce_design}
\end{figure}

Cosmic Explorer is designed as a dual-recycled Fabry-Perot Michelson interferometer, which relies on the differential arm motion for the gravitational-wave readout, as shown in \cref{fig:ce_design}. 
Each arm of the Michelson interferometer consists of two highly reflective mirrors, which serve as test masses and form a Fabry-Perot cavity to increase the power stored in the arms. 
While this enhances the detector's sensitivity to low-frequency signals that vary slowly compared to the storage time of light in the arm cavities, this storage time defines a characteristic frequency, the bandwidth, above which signals are attenuated.
A power recycling mirror placed at the symmetric port forms a power recycling cavity which further increases the power stored in the arms, but has no effect on the shape of the interferometer's response to differential arm motion.

The addition of a signal extraction mirror (SEM) to the antisymmetric port forms a signal extraction cavity (SEC)\footnote{This is also sometimes referred to as a signal recycling cavity (SRC).} which shapes the interferometer's response without decreasing the circulating power in the arms~\citep{Mizuno:1993}. This extraction cavity creates an optical resonance located at\footnote{These equations are valid for third generation detectors and NEMO but require corrections for LIGO due to its more transmissive SEM~\citep{McCuller2021}.}
~\citep{Martynov:2019gvu,McCuller2021}
\begin{equation}
  f_\text{s} = \frac{c}{4 \pi} \sqrt{\frac{T_\text{i}}{L_\text{a} L_\text{s}}}
  = \frac{c}{2}\frac{1}{\sqrt{2\pi\mathcal{F} L_\text{a} L_\text{s}}}
  \label{eq:sec_freq}
\end{equation}
with bandwidth
\begin{equation}
  \gamma_\text{s} = \frac{c T_\text{s}}{8 \pi L_\text{s}},
  \label{eq:sec_bandwidth}
\end{equation}
where $L_\text{a}$ is the length of the arms, $L_\text{s}$ is the length of the SEC, $T_\text{s}$ is the transmissivity of the SEM, $T_\text{i}$ is the transmissivity of the input test masses (ITMs), and $\mathcal{F}=2\pi/T_\text{i}$ is the finesse of the arm cavities. 
In the simplest case where the light is resonant in the SEC, the resulting coupled cavity forms a compound mirror with an effective reflectivity less than that of the ITMs, thus increasing the bandwidth of the interferometer without decreasing the power stored in the arm cavities.

The bandwidth of the extraction cavity resonance described by \cref{eq:sec_freq,eq:sec_bandwidth} is too broad to have an effect in current gravitational wave detectors (it is about \SI{80}{\kilo\Hz} for LIGO), but it can improve the sensitivity centered around $f_\text{s}$ if it is narrowed by creating a ``resonant dip'' in the detector's noise spectral density~\citep{Martynov:2019gvu,Ackley:2020}. Once the parameters of the arms are fixed, the length of the SEC is chosen to target a frequency band of interest according to \cref{eq:sec_freq}, and then the transmissivity of the SEM is chosen to determine the width of this resonance according to \cref{eq:sec_bandwidth}. 
This is the principle behind the NEMO detector's tuning for studying postmerger neutron star physics with a ``long SRC''~\citep{Ackley:2020}, but it is important to note that it is the bandwidth of the resonance\,---\,not the length of the SEC\,---\,that matters. The bandwidth can be narrowed by increasing the reflectivity of the SEM or by increasing the length of the SEC. Cosmic Explorer's long arms require both a relatively short SEC to target postmerger gravitational waves combined with a more highly reflective SEM to narrow the bandwidth. Lowering the transmissivity of the SEM broadens the bandwidth of this resonance, removing the resonant dip in the noise, and improves the low and midband frequencies.

To be able to modify the tuning \textit{between} observing runs to target different science goals, the required changes to the detector must be minimal in practice. For this reason, we assume that the arm cavities and the length of the SEC are constant, and that only the SEM can be switched between observing runs\,---\,a relatively straightforward change.\footnote{One of the filter cavity mirrors would also need to be switched, but this is again a one-time straightforward change which adds no constraints to the tuning considerations discussed here.} Note that this means the location of the resonant dip in the sensitivity $f_\text{s}$ is fixed by the infrastructure; only its width $\gamma_\text{s}$ can be tuned. Additionally, unlike the costly \SI{320}{\kg} Cosmic Explorer test masses, the SEM is a smaller optic that can be acquired and switched at minimal cost and effort. There are several design considerations in choosing the arm cavity finesse $\mathcal{F}$ which have secondary implications for how effectively the detector can be tuned in practice and are discussed in \cref{sec:technology_drivers}. We emphasize that we are not proposing to dynamically tune the location of the SEC resonance to track an inspiral signal as proposed in, for example, \citep{Meers1993,PhysRevD.90.102003}.

Finally, we emphasize that there is no \textit{microscopic} detuning of the signal extraction cavity length, on the scale of the laser wavelength, leading to the characteristic ``double dip'' caused by an optical spring and an optical resonance~\citep{BnC}. In such schemes, only one of the signal sidebands is resonant in the SEC. In the scheme described here, as well as in \citep{Ackley:2020,Martynov:2019gvu}, there is no microscopic detuning of the SEC and therefore both sidebands are resonant in the SEC and there is no optical spring. The choice of $L_\text{s}$ determining the SEC resonance and bandwidth is a \textit{macroscopic} one.

Since the sensitivity is degraded near the free spectral range $f_\text{fsr}=c/2L_\text{a}$ \citep{Essick:2017wyl}, increasing the arm length beyond \SI{40}{\km}, for which $f_\text{fsr}\approx \SI{3.7}{\kilo\Hz}$, is not constructive. This motivates our consideration of a \SI{20}{\km} detector with a correspondingly higher $f_\text{fsr}\approx\SI{7.5}{\kilo\Hz}$, which has better high-frequency sensitivity at the expense of worse broadband sensitivity.

\subsection{Network of Gravitational-wave Detectors and Cosmic Explorer Configurations} \label{subsec:networks}

\begin{figure}
  \centering
  \includegraphics[width=\columnwidth]{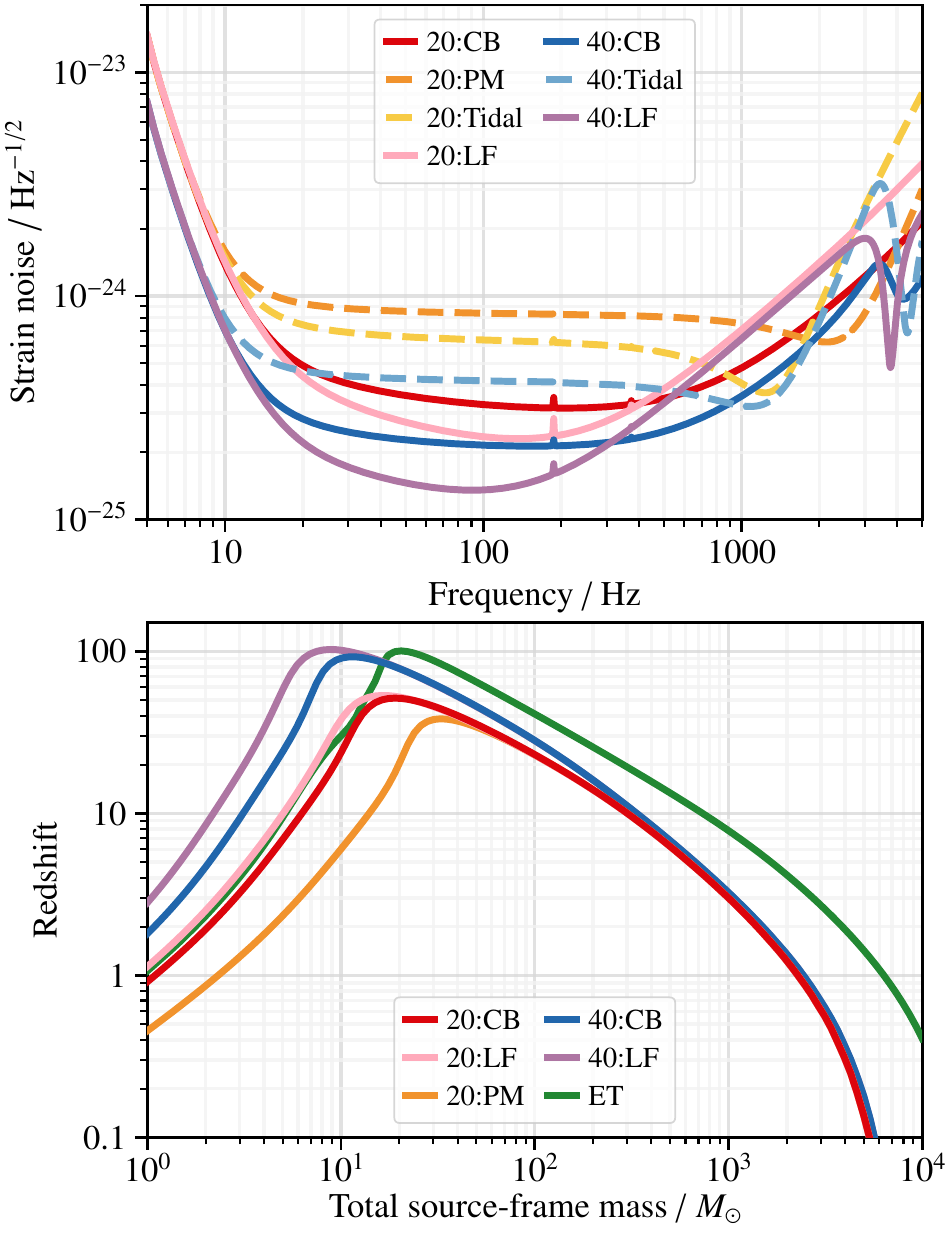}
  \caption{The top plot summarizes the strain sensitivity of the tuned configurations of interest of the \SI{20}{\km} and \SI{40}{\km} Cosmic Explorer (CE) detectors. 
  The compact binary (CB) configuration is the design sensitivity of respective (\SI{40}{\km} or \SI{20}{\km}) observatory.
  Each detector can be tuned to observe with a high-frequency optimized sensitivity -- postmerger optimized (PM) or binary neutron-star tidal (Tidal), and a low-frequency optimized sensitivity (LF).
  The bottom plot shows the horizon redshift as a function of total mass (equal component mass binary) for the corresponding detector configuration along with Einstein Telescope (ET).} 
  \label{fig:4020net}
\end{figure}

We consider the Cosmic Explorer observatory in a background of different plausible gravitational-wave detector networks. However, to underscore the importance of the Cosmic Explorer detectors we also consider networks in its absence. These networks are summarized below:
\begin{itemize}
        \item \textit{Second generation or 2G} network that assumes aLIGO Hanford, Livingston and India are observing at A+ sensitivity. Advanced Virgo and KAGRA at their design sensitivity.
        \item \textit{Voyager} network that assumes aLIGO Hanford, Livingston and India are observing at Voyager sensitivity. Advanced Virgo and KAGRA at their design sensitivity.
        \item \textit{Tuned Voyager India} network that assumes aLIGO Hanford and Livingston are observing at Voyager sensitivity. LIGO India is observing in a post-merger optimized configuration. Advanced Virgo and KAGRA at their design sensitivity.
        \item \textit{Voyager+ET} network that assumes aLIGO Hanford, Livingston and India are observing at Voyager sensitivity. Advanced Virgo and KAGRA at their design sensitivity. ET is operating at it's design sensitivity.
\end{itemize}

\begin{deluxetable}{lcc}
\tablecaption{Cosmic Explorer parameters for the configurations discussed here. It is possible to switch between the compact binary, low-frequency, and one of the two high-frequency tunings (either post-merger or tidal) by changing the signal extraction mirror. However, it is not possible to switch between the two high-frequency configurations as the signal extraction cavity length needs to be changed as well. \label{tab:ce_params}}
\tablewidth{7000pt}
\tablehead{\colhead{Parameters~~~~~~} & \colhead{~~~~~\SI{40}{\km} CE~~~~~}  & \colhead{~~~~~\SI{20}{\km} CE}~~~~~}
\startdata
$L_{\text{a}}$ & \SI{40}{\km} & \SI{20}{\km} \\
Arm Power & \SI{1.5}{\MW} & \SI{1.5}{\MW} \\
$\mathcal{F}$ & 450 & 450 \\
SEC losses & \SI{500}{\ppm} & \SI{500}{\ppm} \\
\tableline 
\multicolumn{3}{c}{Compact binary and Post-merger} \\
$L_{\text{s}}$ & \SI{20}{\m} & \SI{34}{\m} \\
$T_{\text{s}}$ (CB) & \SI{0.02}{\nothing} & \SI{0.04}{\nothing} \\
$T_{\text{s}}$ (PM) & - &  \SI{4.5e-3}{\nothing} \\
$T_{\text{s}}$ (LF) & \SI{0.125}{\nothing} & \SI{0.15}{\nothing} \\
\tableline 
\multicolumn{3}{c}{Compact binary', Tidal and Low-Frequency} \\
$L_{\text{s}}$ & \SI{60}{\m} & \SI{190}{\m} \\
$T_{\text{s}}$ (CB') & \SI{0.02}{\nothing} & \SI{0.04}{\nothing} \\
$T_{\text{s}}$ (Tidal) & \SI{4.5e-3}{\nothing} & \SI{8e-3}{\nothing} \\
$T_{\text{s}}$ (LF') & \SI{0.125}{\nothing} & \SI{0.15}{\nothing} \\
\enddata
\tablecomments{CB' and LF' are the compact binary and low-frequency configurations with the alternate Cosmic Explorer infrastructure used for the tidal tuning. These configurations have similar sensitivities to CB and LF, but are not considered explicitly in this paper.}
\end{deluxetable}

As stated earlier, each Cosmic Explorer detector can operate in three different configurations that are tuned for either low frequencies~(LF, \cref{sec:low_freq_tuning}), compact binary signals~(CB, the nominal broadband tuning), or high frequency signals---either post-merger~(PM, \cref{subsec:pmr_tune}) or tidal (\cref{subsec:tidal_tune}).
We explore design options for Cosmic Explorer facilities with arm-lengths of 10, 20, 30, and \SI{40}{\km}. For simplicity, we will focus on the design with a baseline arm length of \SI{20}{\km} and \SI{40}{\km}. The various tuned Cosmic Explorer sensitivities are labeled as follows
\begin{itemize}
        \item \textit{20:CB or 40:CB} represents a \SI{20}{\km} or a \SI{40}{\km} Cosmic Explorer detector, respectively, observing for compact binary. This configuration serves as a baseline configuration to compare improvements from tuning.
        \item \textit{20:PM} represents a \SI{20}{\km} Cosmic Explorer detector which is optimized for post-merger oscillations. The \SI{40}{\km} post-merger is not considered as it has marginal improvement in post-merger sensitivity due to the reduced sensitivity at the $f_\text{fsr}\approx \SI{3.7}{\kilo\Hz}$.
        \item \textit{20:Tidal or 40:Tidal} represents a \SI{20}{\km} or a \SI{40}{\km} Cosmic Explorer detector, respectively, observing with an improved sensitivity for measuring the tidal effects in binary neutron-star mergers.
        \item \textit{20:LF or 40:LF} represents a \SI{20}{\km} or a \SI{40}{\km} Cosmic Explorer detector, respectively, observing in low-frequency optimized configuration.
\end{itemize}
The corresponding spectra for the \SI{20}{\km} and \SI{40}{\km} detectors along with the tunable configurations is summarized in \cref{fig:4020net} while the parameters are summarized in Table~\ref{tab:ce_params}.

We consider these Cosmic Explorer observatories in a background \textit{2G} network, \textit{Einstein Telescope} (ET) network, and \textit{Cosmic Explorer South} (CES) observatory. Cosmic Explorer South is assumed to be a \SI{20}{\km} post-merger optimized detector.

\section{High Frequency Configurations}
\label{sec:high_freq_tuning}

\subsection{Post-merger signal}
\label{sec:pmr_signal}

The remnant of a binary neutron star merger is hot, dense and rapidly rotating. Depending on mass, spin, magnetic field strength, the unknown equation of state of dense matter at finite temperature, and the processes of neutrino emission, the remnant may collapse immediately to a black hole or remain as a (meta-)stable neutron star supported by uniform or differential rotation (see \citet{DietrichHinderer2021} for a recent review). In the latter case, merger-induced oscillations of the remnant produce post-merger gravitational waves. These gravitational waves have a complex frequency-domain morphology. A characteristic peak frequency is attributable to the fundamental quadrupole oscillation mode~\citep{BausweinStergioulas2016,BausweinJanka2012,bauswein2012equation}, and secondary frequency-domain peaks are due to transient non-axisymmetric deformations and the interaction between quadrupole and quasi-radial modes~\citep{BausweinStergioulas2015}. 
The amplitude and duration of the post-merger emission are particularly sensitive to processes involving magnetic field amplification and neutrino production \citep{SarinLasky2021}.

\begin{figure}
  \centering
  \includegraphics[width=\columnwidth]{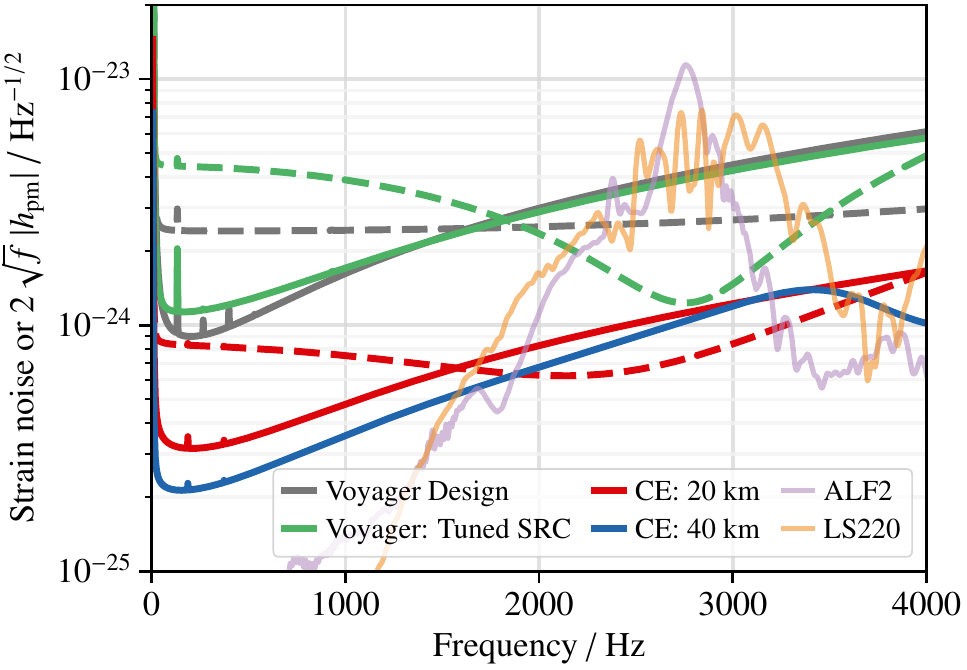}
  \caption{The solid lines shows the strain sensitivity of the corresponding detector in compact binary and the dashed line shows the post-merger tuned configuration. 
  The purple and the orange traces show the sky-averaged source-averaged spectrum of the post-merger signal at \SI{200}{\mega\parsec} for the equation of states ALF2 and the LS220, respectively.
  These traces highlight the post-merger signal of the population of neutron stars extends over a wide range of frequencies, which are equation of state dependent.
  Proposed narrow-band configurations with a bandwidth of few tens hertz like detuned signal recycling cavity are therefore of limited applications for the observation of the post-merger signal, specially when the chances of observation of post-merger signals with a high SNR ($>$8) is low.
  }
  \label{fig:pmr_strain}
\end{figure}

Observations of post-merger signals from a population of binary neutron star mergers can probe finite-temperature matter across the density scale realized in hypermassive remnants. Joint pre- and post-merger gravitational wave observations are especially valuable as  a potential tracer of hadron-quark phase transitions at supranuclear densities~\citep{BausweinBastian2019,MostPapenfort2019}. Post-merger spectra averaged over the whole-sky and source population of the binary neutron star mergers are overplotted in \cref{fig:pmr_strain} for two choices of equation of state.

The sensitivity of gravitational-wave detectors to the post-merger signal has been studied in the context of Advanced LIGO and Virgo~\citep{ClarkBauswein2016,Torres-RivasChatziioannou2019}. Only the loudest binary neutron-star mergers are expected to yield detectable post-merger gravitational radiation for second-generation gravitational-wave detectors. The prospects for third-generation gravitational-wave detectors are more optimistic. In this study, we use post-merger waveforms from the CoRe database of numerical simulations of binary neutron star mergers~\citep{DietrichRadice2018}. The simulations span 164 distinct binaries, with total masses ranging from \SIrange{2.4}{3.5}{\solarmass}, and 17 different equations of state.
In the time-series waveform of the inspiral, merger, ringdown, and post-merger of binary neutron stars, the post-merger oscillations of the remnant are defined after the amplitude of the ringdown has damped down to zero (or a numerical minimum). The post-merger SNR is then defined for the post-merger-only part of the waveform $h_{\text{pm}}$ according to
\begin{equation}
  \mathrm{SNR}^2_\text{pm} = 4\, \re \int_{f_\text{min}}^{f_\text{max}} \frac{\tilde{h}_\text{pm}(f) \tilde{h}_\text{pm}^*(f)}{S(f)}\, \mr{d}f
\end{equation}
where we integrate the post-merger part of the waveform from $f_\text{min} = \SI{1}{\kHz}$ to $f_\text{max} = \SI{4}{kHz}$ to calculate the post-merger SNR, $\tilde{h}_\text{pm}$ represents the Fourier transform of $h_\text{pm}$, $^*$ denotes the complex conjugate, and $S(f)$ is the detector noise spectrum~\citep{ReadBaiotti2013}.

\subsection{Post-merger Tuning}\label{subsec:pmr_tune}
\begin{figure*}
  \centering
  \includegraphics[width=0.95\textwidth]{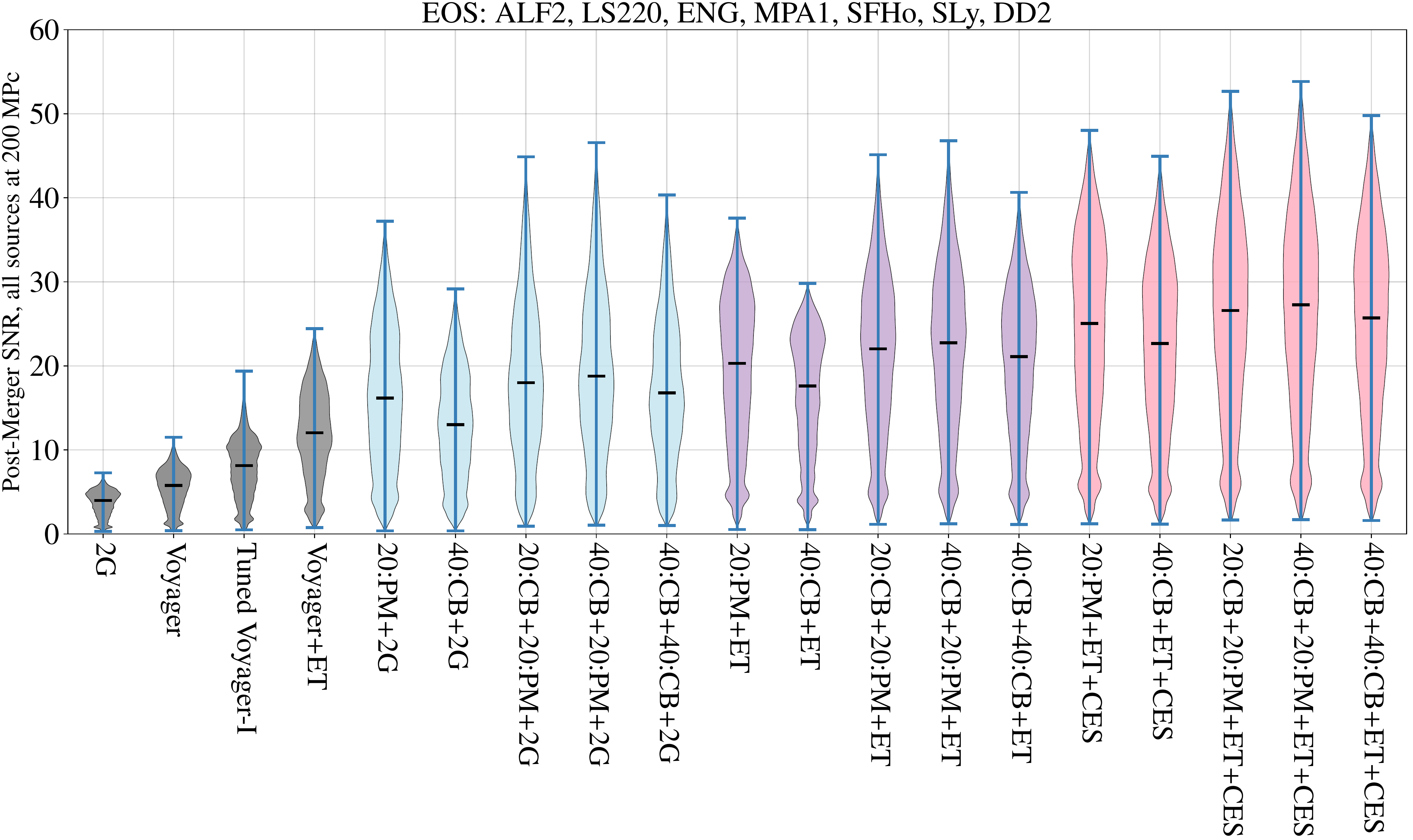}
  \caption{Sky-averaged, source-averaged, EoS-averaged (see \cref{sec:pmr_signal}) SNR of the post-merger signal for \SI{160}{\kilo\nothing} binary neutron star sources at \SI{200}{\mega\pc}. 
  The performance of networks in the absence of Cosmic Explorer detectors are shown in gray, Cosmic Explorer in a background of 2G networks are shown in blue, networks with Cosmic Explorer and Einstein Telescope are shown in purple, and networks with Cosmic Explorer, Einstein Telescope, and an additional Cosmic Explorer South observatory are shown in pink. 
  \cref{subsec:networks} defines all the networks compared above.
  }
  \label{fig:ce_violin}
\end{figure*}

We consider two stages of post-merger optimized tuning. 
First, tuning of the proposed CE design with respect to current equation of state constraints~\citep{GW170817,GW170817_EoS,GW170817_Joint,De:2018uhw,Radice_2018,radice2019multimessenger,Carson_2019_EoS,capano2020stringent,MillerLamb2019,RileyWatts2019,MillerLamb2021,RileyWatts2021}.
Second, a more aggressive tuning based on potential future improvements in our knowledge of the post-merger frequencies as the constraints on the equation of state improve, which we simulate by fixing the equation of state.

We also explore post-merger tuning possibilities for gravitational-wave detectors for the proposed Voyager upgrade to the current \SI{4}{\km} LIGO facilities. 
For aLIGO Hanford and Livingston we only consider changing the transmissivity of the signal extraction mirror. 
However, as the aLIGO India facility is still under construction, we allow the length of the signal extraction cavity to change as well. We refer to this post-merger optimized configuration as \textit{Tuned Voyager India}. Note this does not affect the optimal broadband sensitivity of the LIGO India facility but allows the possibility for it to operate in a high-frequency tuned configuration.

The length of the signal extraction cavity (SEC) and transmissivity of the signal extraction mirror (SEM) are optimized for the 20 and \SI{40}{\km} Cosmic Explorer detectors by maximizing the SNR of a constant post-merger strain of \SI{1e-25}{/\,Hz^{1/2}} from \SI{2}{\kilo\Hz} to \SI{4}{\kilo\Hz}. The Cosmic Explorer facility is built with this optimal SEC length for post-merger signals and the SEM transmissivities are then changed to switch between the compact binary and low-frequency tunings.

As the observation of the post-merger oscillation signal will be limited to nearby sources, it is critical to ensure the post-merger tuned configuration is optimal for a population of sources. 
Given the low astrophysical rate, the narrow-band configurations increase the risk of missing the post-merger signal completely, owing to the uncertainty in the equation of state and the source parameters.
To quantify the prospects of observation of post-merger oscillations in the above networks, we marginalize over the plausible equation of states in the CoRe database with a broad range of component masses of binary neutron stars
\citep{GW170817,GW170817_EoS,GW170817_Joint,De:2018uhw,Radice_2018,radice2019multimessenger,Carson_2019_EoS,capano2020stringent,MillerLamb2019,RileyWatts2019,MillerLamb2021,RileyWatts2021}.
For all of the plausible equation of states, the post-merger signal is injected across the sky at fixed distances of \SI{100}{\mega\pc}, \SI{200}{\mega\pc}, \SI{500}{\mega\pc} and \SI{1}{\giga\pc}.
The frequency shift of the post-merger signal due to the cosmological redshift is considered.
The post-merger signal is then projected on the different detectors considered in the study; see \cref{subsec:networks}.
The post-merger SNR of the network is calculated by the quadrature sum of the SNRs in each detector.
\cref{fig:ce_violin} summarizes the post-merger SNR for approximately \SI{160}{\kilo\nothing} injections at \SI{200}{\mega\pc} averaged over the different equations of state. We find that the \SI{20}{\km} post-merger optimized Cosmic Explorer offers the loudest post-merger SNRs across all plausible neutron star equations of state. 
It is important to note that the non-observation of post-merger signals with third-generation gravitational-wave detectors will hint at softer equations of state, which do not support post-merger oscillations of the hypermassive remnant.

To constrain the hot equation of state and observe phase transitions in the post-merger remnant requires multiple observations of binary neutron star mergers across the mass spectrum. 
Using the median of the sky-averaged, equation of state-averaged, post-merger SNRs, and an observed merger rate of \SI{320}{\giga\pc^{-3}\yr^{-1}}~\citep{GWTC1,GWTC2}, we find that a \SI{40}{\km} Cosmic Explorer in a background of 2G networks will detect 40 events per year with a post-merger SNR greater than 8. Two \SI{40}{\km} Cosmic Explorer detectors can observe 80 such events. A single \SI{20}{\km} post-merger optimized detector can observe 80 such sources each year while a network of a \SI{40}{\km} and a \SI{20}{\km} Cosmic Explorer can observe 120 post-merger signals each year. Each of these signals can then be coherently combined to constrain the neutron star equation of state, and facilitate the understanding of hot, dense matter~\citep{bose2017, Tsang2019}

One may wish to revisit the post-merger tuning \textit{if}, over the next decade, constraints on the neutron star equation of state improve \textit{prior} to the construction of Cosmic Explorer. However, significant improvement from the proposed post-merger tuning will be limited for two reasons.
First, any improvements coming from narrowing the bandwidth of the high frequency dip are equation of state dependent.
As an example of this scenario, we choose two equations of state from the CoRe database that sample the population of binary neutron stars---ALF2 and LS220. We inject each of these numerical waveforms/sources assuming an isotropic distribution across the sky at \SI{200}{\mega\parsec}. The corresponding strains are then averaged, which allows one to access the frequencies of interest for the population of sources for the particular equation of state; see \cref{fig:pmr_strain}. For ALF2, the bandwidth of the \SI{20}{\km} post-merger optimized CE can be further tuned to provide an improvement but further improvements in post-merger SNR from bandwidth tuning are limited for LS220. 
This is because the post-merger signal of LS220 spans a wide frequency band. Any narrow-band configuration will therefore be non-optimal.
Second, even if it were beneficial to narrow the bandwidth, it would be technically extremely challenging to do so due to loss in the signal extraction cavity; see \cref{sec:technology_drivers} for a detailed discussion.

\subsection{BNS Tidal Effects}\label{subsec:tidal_tune}
Instead of the post-merger tuning, the Cosmic Explorer detectors can be tuned to improve the measurement of neutron-star tidal parameters, which are dependent on the cold equation of state.
The non-zero tidal deformability of the neutron stars in a compact binary coalescence changes the gravitational-wave phase accumulated over hundreds of cycles during the inspiral. 
The measurability of tidal effects is facilitated by improved sensitivity at higher frequencies, close to the contact frequency of the merger.
We find Cosmic Explorer configurations tuned to target the late inspiral up to the contact frequency for a population of binary neutron stars using a similar analysis as that used to find the post-merger tunings discussed in \cref{subsec:pmr_tune} using a phenomenological waveform with strain proportional to frequency between \SI{500}{\Hz} and \SI{1500}{\Hz}.

We quantify the benefits of specific configurations using the integrated measurability of tidal effects in the gravitational-wave signal. 
The measurability per unit frequency of tidal effects is proportional to 
$f/S(f)$, where $S(f)$ is the power spectral density of the detector~\citep{Damour:2012yf}. 
We approximate the relative measurability of tidal effects in binary neutron mergers for different configurations of Cosmic Explorer by this integral.
We integrate from 10 Hz up to the contact frequency, $C_f$, of the binary neutron star system. 
Hence, we define the tidal measurability, $M_\Lambda$, as
\begin{equation}\label{eq:meas_int}
M_\Lambda = \int_{\SI{10}{\Hz}}^{C_f} \frac{f}{S(f)}\,\mr{d}f
\end{equation}
The contact frequency is used as the upper bound of integration to separate this measurement from post-merger measurements. 

As the measurability function is independent of the equation of state, the ratio of the tidal measurability for two detector configurations is only a function of the detector noise spectrum and this contact frequency, which determines the bounds of integration in \cref{eq:meas_int}.
We use this relationship to compare the tidal measurability of a variety of systems with different Cosmic Explorer configurations.

\begin{figure}
  \centering
  \includegraphics[width=\columnwidth]{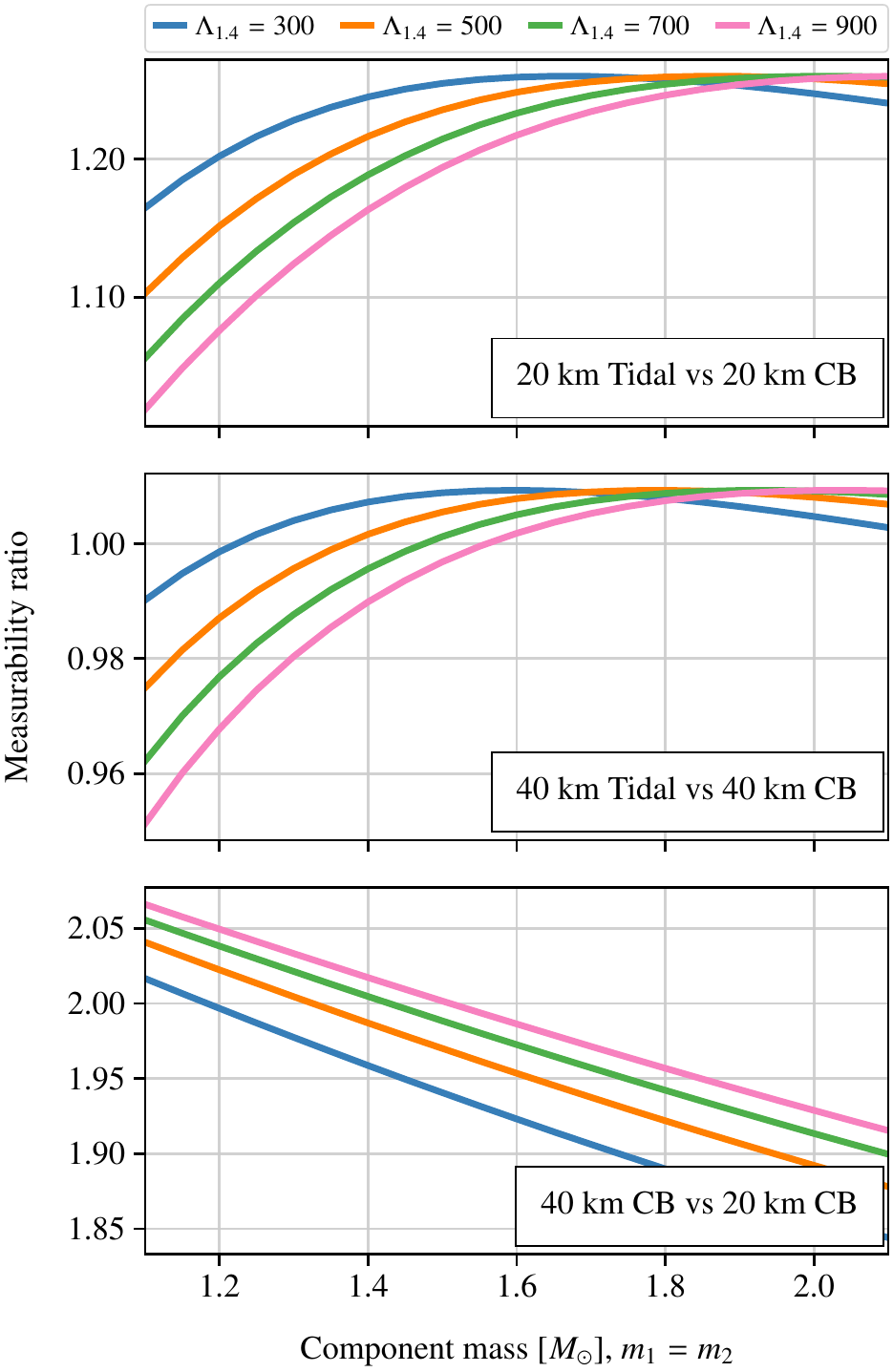}
  \caption{Ratio of measurable tidal information for a Cosmic Explorer detector tuned to tidal measurements versus a broadband configuration. 
  In both the \SI{20}{\km} and \SI{40}{\km} case, a detector can be tuned to improve the measurability of the tidal deformability.
  Comparing a \SI{20}{km} and \SI{40}{\km} facility, the \SI{40}{\km} case would significantly increase the overall tidal measurability, although the additional benefits for a \SI{40}{km} tidal configuration are reduced.
  In fact, a \SI{40}{\km} broadband configuration would measure tidal effects better than a \SI{20}{\km} tidal detector for all equations of state and masses considered.}
  \label{fig:tidal_sensitivity}
\end{figure}

The contact frequency of a binary system with two equal mass neutron stars of mass $m$ and radius $R(m)$ is given by~\citep{De:2018uhw}
\begin{equation}
    C_f(m) = \left(\SI{1530}{\Hz}\right) \left(\frac{m}{\SI{1.4}{\solarmass}}\right)^{1/2} \left(\frac{R(m)}{\SI{12.62}{\km}}\right)^{-2/3}.
\end{equation}

To approximate the contact frequency over a wide range of masses, we assume that all neutron stars have a constant radius, determined by the radius of a 1.4 solar mass neutron star with tidal deformability $\Lambda_{1.4}$. 
This common radius $R_{1.4}$ is~\citep{De:2018uhw}
\begin{equation}
    R(m) \approx R_{1.4} = \left(\SI{12.62}{\km}\right)\left(\frac{\Lambda_{1.4}}{500}\right)^{1/6}.
\end{equation}

To verify that this calculation of $M_\Lambda$ holds for state-of-the-art waveforms, we have compared this analytic approximation to the measurability per unit frequency of \textsc{IMRPhenomDNRTidal} waveforms~\citep{Dietrich:2018uni,Dietrich:2017aum}. 
This is done computationally by calculating the gradient of the match versus frequency for two waveforms with similar tidal deformability values. 
We confirm that the analytic result holds up until the contact frequency for the range of masses and tidal deformability values explored in this work.

\Cref{fig:tidal_sensitivity} shows the results of this comparison for both 40 km and 20 km tidal configurations of Cosmic Explorer. 
Over a range of masses and values for $\Lambda_{1.4}$, a tidal configuration for a 20 km Cosmic Explorer increases the tidal sensitivity by up to \SI{25}{\%}, 
while a tidal configuration for a \SI{40}{\km} Cosmic Explorer only increases the tidal measurability by up to \SI{10}{\%}. 
The mass for which the tidal measurability ratio is maximized at a fixed value of $\Lambda_{1.4}$ is determined by the frequency range where the detector sensitivity is maximized by the tidal configuration. 
The optimal configuration is such that the maximal sensitivity peak is set to just below the expected contact frequency.
In practice, the exact tidal configuration can be set to the optimal configuration based on tidal information already known from second generation observations and the mass range of interest. 
Comparing between different facilities, a \SI{40}{\km} broadband detector increases the tidal measurability by up to 
\SI{105}{\%} compared to a \SI{20}{\km} broadband detector. 

\section{Low Frequency Configuration}
\label{sec:low_freq_tuning}

In this section, we motivate low frequency tuned configurations focused on improving the detection probability of the binary-black hole population at high redshifts, such as from the remnants of POP-III stars~\citep{Woosley_1995,Ostriker_1996,Bromm_2002,Heger_2002,Schaerer_2002} and seed black-hole binaries~\citep{Ferrarese_2000,Kormendy_2013,Greene_2020}. 
These populations are at high redshift and are comprised of heavier binaries~\citep{PhysRevLett.116.221101,PhysRevD.100.104036,PhysRevD.103.122002}, which limits their predominant gravitational-wave signals to below \SI{50}{\Hz}~\citep{Ng_2021}. 
Low frequency improvements facilitate improved tests of General Relativity. 
A key test of General Relativity is to precisely measure the amplitude and frequency (as a function of mass and spin) of the quasinormal modes of black hole remnants~\citep{berti2018extreme1,berti2018extreme2,Cardoso_2019}. 
The loudest sources observed with third-generation gravitational-wave detectors will provide the most stringent tests of General Relativity. 
Thus, we will focus on sources which are close and we can safely assume that this binary population is comprised of stellar mass black-holes.
The low frequency tuning is complimentary to the high frequency tuning and can be realized by switching the reflectivity of the signal extraction mirror.

At a fixed distance, heavier mass binaries have higher gravitational-wave amplitudes, and the remnant has lower  quasinormal frequencies.
The low frequency tuned configurations are tuned by maximizing the SNR using a phenomenological frequency-domain waveform of the fundamental ringdown of stellar mass black holes.
We use astrophysically weighted populations to construct the phenomenological waveform~\citep{Srivastava_supernovae}.
We will quantify the performance of the low frequency tuned configuration in the next sections.

\subsection{SNR Improvements from low-frequency tuning}\label{subsec:snr_imp}
\begin{figure}
  \centering
  \includegraphics[width=\columnwidth]{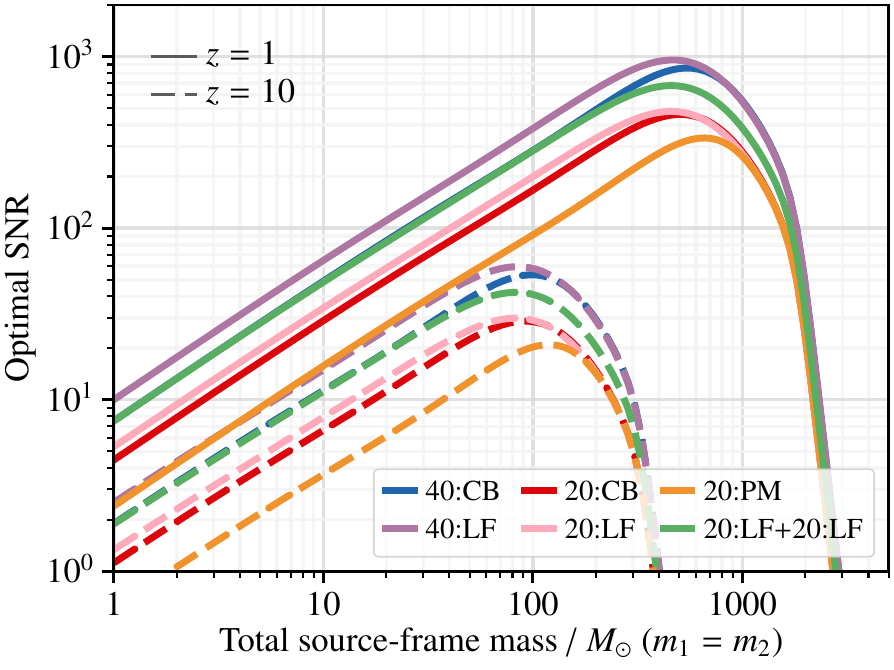}
  \caption{The optimal SNRs as a function of the total mass of the binary with two equal mass components.
  A \SI{20}{\km} low frequency tuned detector provides up to \SI{15}{\%} improvement in SNR relative to the broadband configuration while a low frequency tuned \SI{40}{\km} detector provides a \SI{30}{\%} improvement.}
  \label{fig:ce_lf_snrs}
\end{figure}

To quantify the performance of low frequency tuned configurations, we compute the optimal SNR of equal mass binaries, with total mass ranging from \SI{1}{\solarmass} to \SI{e4}{\solarmass}.
This population is considered at different distances (or redshift) to quantify the effects of the cosmological redshift of the gravitational-wave spectrum in the detector frame.
The \cref{fig:ce_lf_snrs} summarizes the comparisons between the broadband and the low-frequency tuned configuration.
We find that for a large number of the binaries, a \SI{40}{\km} low frequency tuned configuration provides up to \SI{30}{\%} improvement from the broadband detector. 
A \SI{20}{\km} low frequency tuned detector provides up to \SI{15}{\%} improvement relative to the broadband configuration.
These improvements are achieved even for sources at high redshifts. 
In particular, for heavier compact binaries at a redshift of 10 or higher, like POP-III star population, this improvement in SNR will improve the detection prospects. 

\begin{figure}
  \centering
  \includegraphics[width=\columnwidth]{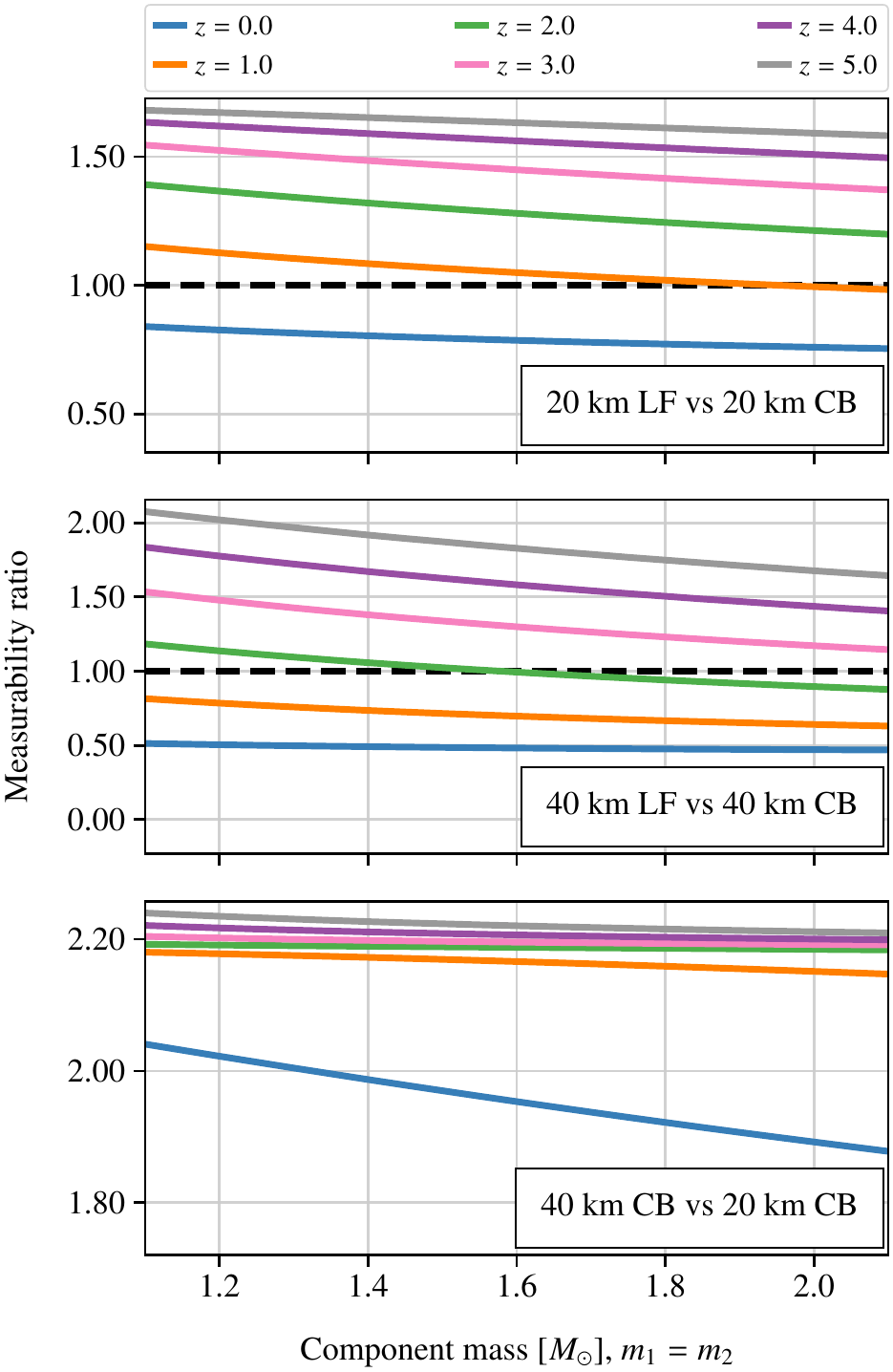}
  \caption{Ratio of measurable tidal information for a Cosmic Explorer detector tuned to low frequencies versus a broadband configuration for different redshifts. 
  Also shown is this ratio for a \SI{40}{\km} facility in a broadband configuration versus a \SI{20}{\km} facility also in a broadband configuration. 
  In all cases, $\Lambda_{1.4}$ is assumed to be 500.
  For redshifts greater than 1--2, the amount of tidal information available is greater 
  with a low frequency configuration than with a broadband configuration and a \SI{40}{\km} facility is better than a \SI{20}{\km} facility.} 
  \label{fig:redshift_tidal}
\end{figure}

\subsection{Continuous Waves}\label{subsec:cw}
Neutron stars with rotational frequencies in the audio band could be a source of continuous gravitational waves for Cosmic Explorer, lasting for millions of years. Neutron stars that are perfectly spherically symmetric or are spinning about their symmetry axis emit no radiation since their quadrupole would not vary with time. However, non-axisymmetric neutron stars emit gravitational waves at twice their spin frequency $f_\text{GW}=2f_\text{spin}$. Their amplitude depends on the ellipticity $\epsilon \equiv (I_{xx}-I_{yy})/I_{zz},$ where $I_{kk},$ $k=x,y,z,$ are the principal moments-of-inertia with respect to the rotation axis. For a neutron star at a distance $D$ the amplitude is
\begin{equation}
h \sim \frac{4\pi^2G}{c^4 D}\, \epsilon\,I_{zz}\,f_{\text{GW}}^2
\end{equation}
Typical neutron star moments are $I_{zz}\sim \SI{3e38}{\kg.\m^2}.$ 
The Crab pulsar (B0531+21) with a spin frequency of \SI{30}{\Hz} (gravitational-wave frequency of \SI{60}{\Hz}), located at a distance of \SI{2}{\kilo\pc}, will have an amplitude of $h \simeq \num{5.7e-29}$ if its ellipticity is $\epsilon \sim \num{e-8}.$ The only way to find such signals is to matched filter the data over a year accumulating billions of wave cycles in the Fourier transform of the data demodulated to account for Earth's rotation and revolution and pulsar's spin down.  Indeed, the signal-to-noise ratio grows as the square-root of the integration period or the number of wave cycles. The characteristic strain amplitude $h_c$ of a signal integrated over a time $T$ is $h_c=h\sqrt{T}$, which for Crab would be $h_c\sim \SI{3e-25}{/\,Hz^{1/2}}(\epsilon/\num{e-8})$ for an integration period of $T=\SI{1}{\yr}$. A millisecond pulsar with a spin frequency of \SI{300}{\Hz} but the same ellipticity will be 100 times louder. 

Signals with characteristic amplitude larger than the amplitude noise spectral density would be detectable with loudness proportional to their height above the noise amplitude. 
A \SI{40}{\km} Cosmic Explorer tuned to lower frequencies (\cref{fig:4020net}, 40 km:LF), will have the best sensitivity to neutron stars of spin frequencies 20--\SI{200}{\Hz}. 
For example, the Crab pulsar would be detectable if its ellipticity was $\epsilon > \num{3e-8}$ after a year's integration. In general, neutron stars of spin frequencies in the range 10--\SI{200}{\Hz} (GW frequencies of 20--\SI{400}{\Hz}) would be accessible to Cosmic Explorer tuned to low frequencies if their ellipticities are larger than about 10 parts per billion or more precisely if
\begin{equation}
    \epsilon \ge \num{e-8} \left ( \frac{f_{\text{GW}}}{\SI{200}{\Hz}}\right )^{-2} \left (\frac{D}{\SI{10}{\kilo\parsec}}\right ) 
    \left (\frac{T}{\SI{1}{\yr}} \right )^{-1/2}.
\end{equation}
The ellipticity is roughly equal to the fractional difference in the size of a neutron star along two principal directions orthogonal to the spin axis. Thus, Cosmic Explorer will be able to constrain fractional difference in the equatorial radii of a millisecond pulsar as small as \SI{100}{\um}.

\subsection{BNS signals from high redshifts}\label{subsec:tides_highz}
While the tidally tuned configurations discussed in \cref{subsec:tidal_tune} are beneficial for measurements of tidal properties in the local universe, such configurations will not simultaneously be optimal for significantly redshifted signals. 
For signals at extremely high redshifts ($ z > 2.0$), an interferometer tuned to low frequencies
will provide a better measurement of tidal parameters. 
To compare the low frequency tuned configuration to a broadband configuration, we use the tidal measurability metric that was introduced in \cref{subsec:tidal_tune}.
The ratio between the tidal measurability of a broadband detector and a detector tuned to low frequencies can be seen in \cref{fig:redshift_tidal}. 
For both the \SI{40}{\km} and \SI{20}{\km} case, a detector tuned to low frequencies will be able to better measure the 
tidal information from high redshift events. 
The redshifting of detector-frame contact frequency for these distant events explains the increase in measurability ratio with respect to redshift. 
Furthermore, a \SI{40}{\km} detector has higher tidal measurability than a \SI{20}{\km} detector from signals at any redshift. 

Although the measurement of a universal nuclear equation of state will be driven by events in the local universe~\citep{Vivanco:2019qnt}, 
measurements of the equation of state at high redshifts will provide additional cosmological information. 
Tidal information from high-redshift events is a potential way to accurately measure
the Hubble constant using only gravitational-wave observations~\citep{Messenger:2011gi}.
Probes of high redshift events will also allow any potential time-evolution of the nuclear equation of state to be measured. 
Any variation in the measured equation of state 
could indicate physics beyond the Standard Model~\citep{Smerechynskyi:2020cfu,Guver:2012ba}. 

\begin{figure*}
  \centering
  \includegraphics[width=0.95\textwidth]{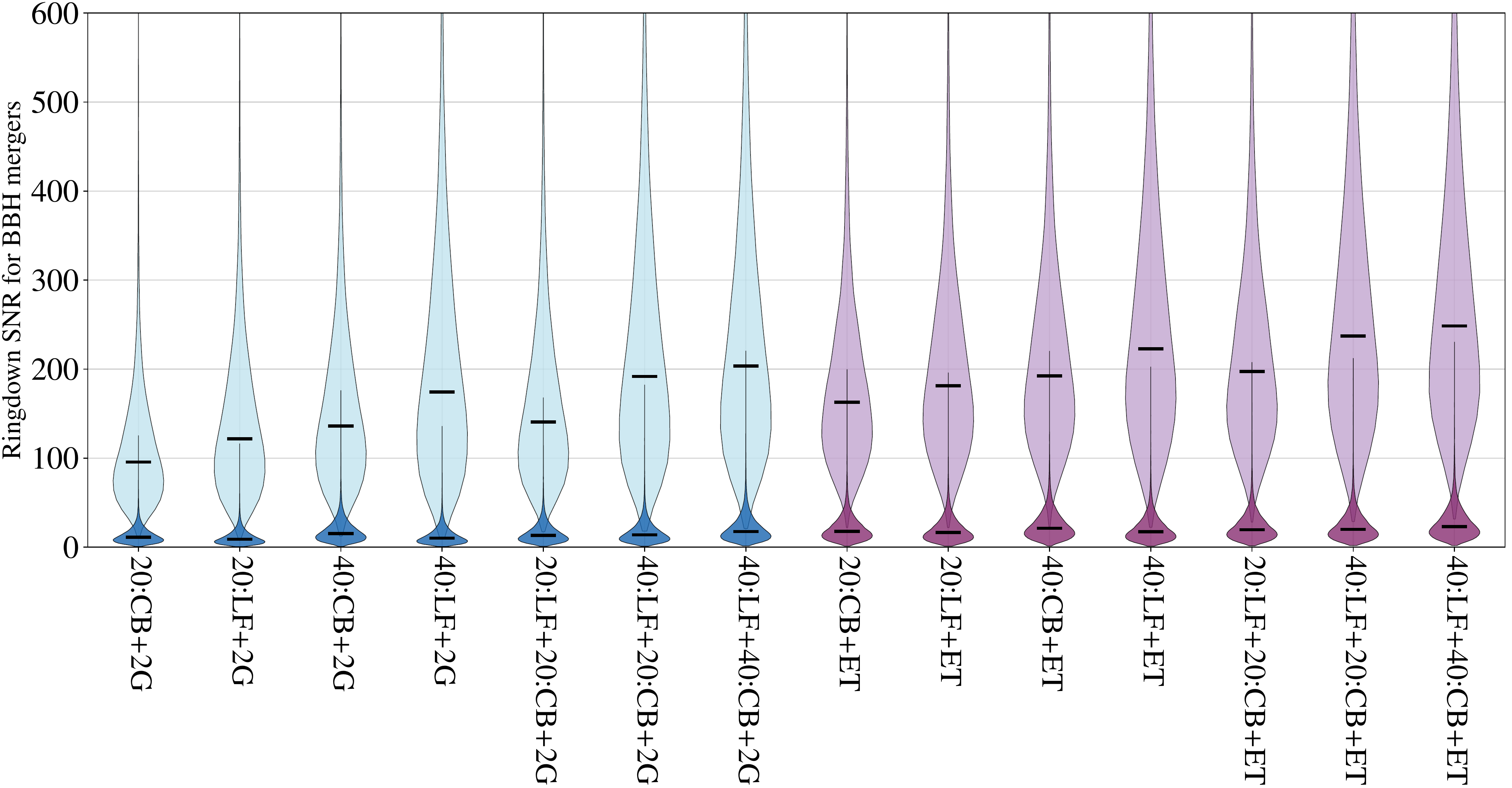}
  \caption{
  Sky-averaged and source-averaged distribution of the ringdown SNR of the 100 loudest events observed with Cosmic Explorer observatories each year (\cref{subsec:grav_tune}) using the observed merger rate of \SI{23.8}{\giga\parsec^{-3}.\year^{-1}} for binary black holes, and  \SI{50}{\kilo\nothing} injections each of low-mass binaries (darker shade) and heavier stellar-mass binaries (lighter shade). The low-frequency optimized configuration improves the observational prospects of the ringdown modes of the remnant black holes. A network of Einstein Telescope and Cosmic Explorer observatories will offer the most stringent tests of General Relativity. \cref{subsec:networks} defines all the networks compared above.}
  \label{fig:ce_violin_bbh}
\end{figure*}

\subsection{Exploring the Nature of Extreme Gravity}\label{subsec:grav_tune}
The population of binary black-holes observed by the aLIGO and Virgo detectors has facilitated key tests of the theory of General Relativity~\citep{PhysRevLett.116.221101,PhysRevD.100.104036,PhysRevD.103.122002}.
These tests of General Relativity include measurement of the consistency of the inspiral-merger-ringdown signal, the spin-induced moments, and polarization of the observed gravitational wave signal.
The measurement of the respective amplitude and frequencies of these quasinormal modes is referred to as black-hole spectroscopy.
Any deviation of the observed spectral features from the predictions of General Relativity will challenge the theory.
We use the SNR of the inspiral-merger-ringdown (discussed in \cref{subsec:snr_imp}), and the ringdown SNR of the remnant black-hole to quantify the performance of Cosmic Explorer configuration to explore the nature of extreme gravity.

The loudest signals during Cosmic Explorer are expected to provide the best tests to General Relativity.
We consider a population of \SI{50}{\kilo\nothing} sources each of lighter and heavier stellar mass binaries. 
The lighter-mass binaries are injected uniformly in component mass between \SI{5}{\solarmass} and \SI{10}{\solarmass}, and the heavier-mass binaries are injected uniformly in mass between \SI{10}{\solarmass} and \SI{70}{\solarmass}.
Both of these source populations are injected uniformly in volume between \SI{100}{\mega\parsec} and \SI{950}{\mega\parsec}.
Using the observed binary black-hole merger rate of \SI{23.8}{\giga\parsec^{-3}\year^{-1}}~\citep{GWTC1,GWTC2},
this corresponds to the 100 loudest sources detectable each year with Cosmic Explorer.
The sky-averaged source-parameter-averaged distribution of the ringdown SNR of the lighter and heavier populations is shown in \cref{fig:ce_violin_bbh}.
In a background network of 2G detectors, we find the median ringdown SNR of the 100 loudest binary black-hole sources with a \SI{20}{\km} Cosmic Explorer is 120 and is 175 with a \SI{40}{\km}. The performance of two \SI{40}{\km} Cosmic Explorer detectors and a network of a \SI{40}{\km} and a \SI{20}{\km} Cosmic Explorer is similar.
The ringdown SNR for the 100 loudest events improves significantly with Einstein Telescope in the network, \cref{fig:ce_violin_bbh}.

\section{Technological drivers and limitations}
\label{sec:technology_drivers}

In this section we summarize how the noise sources and design choices limit the sensitivity of the various detectors and configurations in order to motivate the research and development (R$\&$D) necessary to maximize the scientific output.
It is important to note that most noises are reduced as the arm length is increased and only the quantum noise is affected by the choice of tuning.

\begin{figure}
  \centering
  \includegraphics[width=\columnwidth]{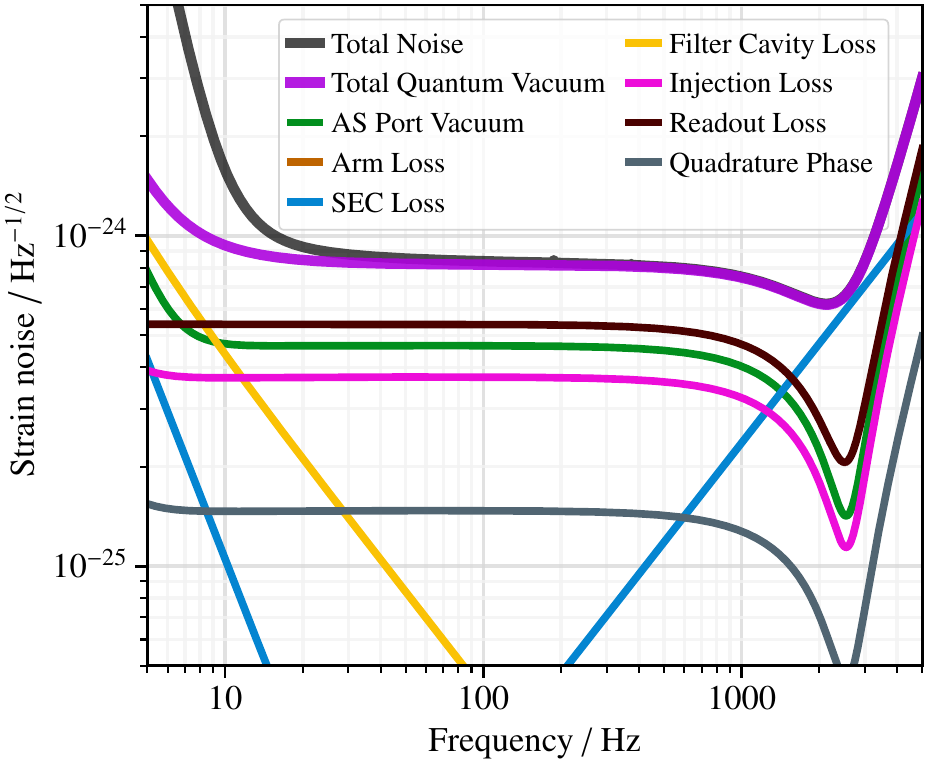}
  \caption{Quantum noise contributions for the \SI{20}{\km} post merger tuned configuration. The total quantum noise (purple) is the sum of the fundamental quantum noise (green) and noises coming from various loss mechanisms and technical noises (other solid colored curves). The black curve is the total noise of the \SI{20}{\km} instrument with an signal extraction loss (SEC) of \SI{500}{\ppm}. The SEC limits the sensitivity and location of the high frequency resonant dip. A much lower SEC loss of \SI{10}{\ppm} is required to suppress its noise contribution below the readout losses in the \SI{20}{\km} postmerger optimized detector. This underscores the need for research into mitigating SEC losses.}
  \label{fig:20km_pm_quantum}
\end{figure}

For all of the post-merger and tidal configurations, the detectors are limited by quantum noise above ${\sim}\SI{20}{\Hz}$. Reducing quantum noise relies on increasing the power stored in the arm cavities, increasing the injected squeezing, and reducing all sources of loss---such as optical, mode-mismatch, scattering, etc.
\cref{fig:20km_pm_quantum} shows the contributions of the various noises to the total quantum noise for the \SI{20}{\km} post-merger configuration, as well as the total noise in black. 
Reducing loss along the input to and output from the main interferometer, including increasing the quantum efficiency of the photodiodes, is essential in achieving the quantum noise targets in the mid-band frequencies.
However, at higher frequencies the losses in the signal extraction cavity (SEC) dominate and limit the sensitivity near the resonant dip for these configurations. SEC loss limits the bandwidth of the compact binary tunings and is not as significant for the low frequency tunings.

Reducing loss in the SEC is thus one of the most critical areas of research needed to realize the high frequency sensitivity goals of Cosmic Explorer.
Since SEC loss is independent of both the tuning and the arm length~\citep{Evans:2021gyd,Miao2019}, it both reduces the sensitivity near the resonant dip \textit{and} limits the frequency to which that dip can be pushed as is illustrated in \cref{fig:20km_pm_quantum}. 
The current noise estimates assume a loss of \SI{500}{\ppm} which includes both optical and mode-mismatch losses discussed below. The corresponding loss in the aLIGO detectors is estimated to be roughly 10 times larger.

The requirements on matching the optical modes between the various optical cavities of the interferometer are likely to be exceedingly strict, and mismatch between these modes is, in some cases, an extra source of SEC loss~\citep{McCuller2021}. Continued development of adaptive mode matching techniques~\citep{PhysRevD.87.082003,Wittel_2014,Brooks2016,PhysRevLett.121.263602,Wittel:18,Cao:20,Srivastava2021} is thus one crucial area of R$\&$D necessary to maximize the high frequency sensitivity.

The quantum noise limiting Cosmic Explorer is inversely proportional to the square root of the arm power, and it is thus important to store high power in the arm cavities. The Cosmic Explorer design calls for \SI{1.5}{\mega\W} arm power, twice as much as the Advanced LIGO design. The arm power in the current detectors has been limited by the presence of particulates in the mirror coatings which absorb the laser power in localized points and thermally distort the mirrors~\citep{Brooks2021,Jia2021,Buikema2020}. Removing, or otherwise compensating the effects of this contamination, is an active area of research. Even if the coatings no longer have this contamination, the power absorbed in the test mass substrates and coatings creates both a thermoelastic deformation of the mirror and a thermally induced lens. All of these effects produce wavefront distortions that require high-resolution wavefront sensing~\citep{Agatsuma:19,Cao:20_PC,PhysRevD.104.042002,Brown:21} and need to be corrected with the adaptive optics discussed above.

Quantum noise is particularly affected by the design of the arm and signal extraction cavities. As shown in \cref{eq:sec_freq,eq:sec_bandwidth}, for a fixed arm length, the arm cavity finesse $\mathcal{F}$, the SEC length $L_\text{s}$, and the signal extraction mirror transmissivity $T_\text{s}$ determine both the location and width of the resonant dip for the post-merger and tidally tuned configurations. Several considerations are important in choosing the finesse. First, SEC loss scales as $\sqrt{\mathcal{F}}$, and so choosing a small finesse directly lowers the high frequency noise. However, indirect effects limit how small $\mathcal{F}$ can be reduced. The power stored in the arm cavities is enhanced by a factor of $\mathcal{F}$. Therefore, for a fixed arm power, increasing $\mathcal{F}$ reduces the power traversing the input test mass and beamsplitter substrates, thus decreasing the power absorbed in these substrates and reducing the associated thermal effects. Furthermore, the coupling of noise from auxiliary degrees of freedom into the gravitational wave signal is suppressed by increasing $\mathcal{F}$. Pending further study, the preliminary Cosmic Explorer design uses the same value of $\mathcal{F}=450$ as does LIGO. Note from \cref{eq:sec_freq} that increasing $\mathcal{F}$ also directly lowers the location of the resonant dip.

With the cavity finesse set, the SEC length determines the location of the resonant dip according to \cref{eq:sec_freq}. The difficulty of matching the optical modes between the SEC and arm cavities, a source of SEC loss as discussed above, is increased as $L_\text{s}$ is decreased. The optimal length is $L_\text{s}=\SI{34}{\m}$ for the \SI{20}{\km} post-merger tuning, which is quite short\,---\,$L_\text{s}=\SI{55}{\m}$ for LIGO and the mode matching problem is more difficult for Cosmic Explorer due to its larger beams required by its longer arms~\citep{PhysRevD.103.023004}. If the length of the SEC needs to be increased, the location of the resonant dip will be decreased with a corresponding reduction in post-merger sensitivity.

The non-quantum noises are not directly affected by the choice of tuning; however, most scale inversely with some power of the arm length~\citep{Abbott2017,Evans:2021gyd}. Indeed, one of the major technical advantages of the Cosmic Explorer design is that much of the increased sensitivity over the second generation detectors, in the mid to high frequencies, comes from increasing the arm length and does not rely significantly on reducing the displacement noises. The \SI{40}{\km} detector is clearly advantageous here and provides a larger margin of error than that of the \SI{20}{\km} detector in the event that some noises do not meet their projected sensitivities.

\begin{figure}
  \centering
  \includegraphics[width=\columnwidth]{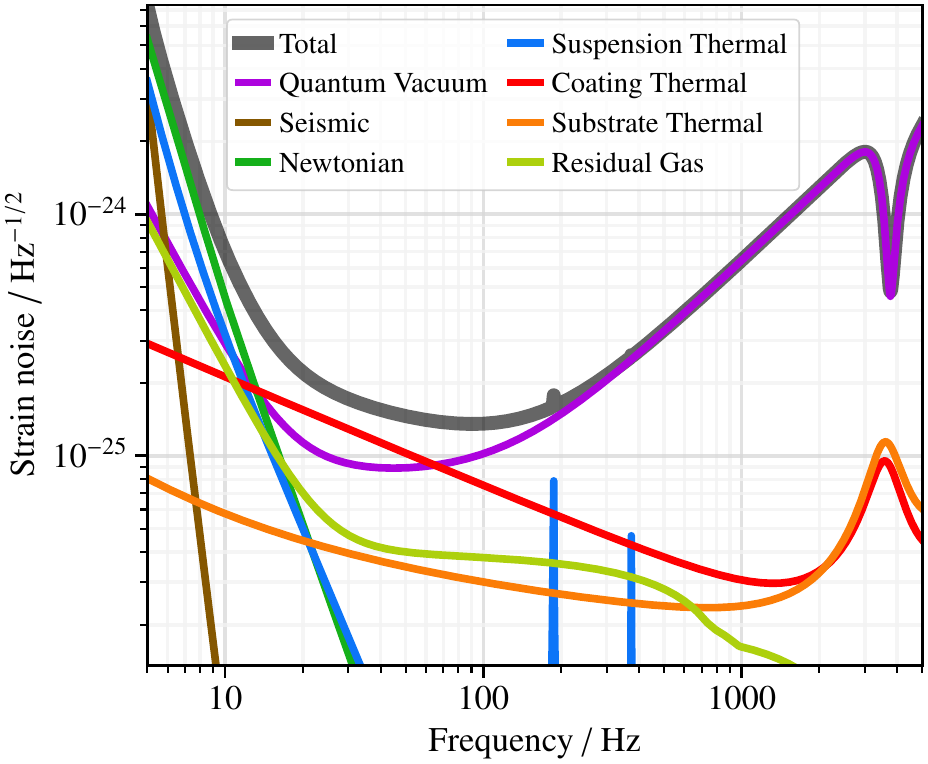}
  \caption{The noise budget of the low frequency tuned \SI{40}{\km} detector. We note that the low frequency sensitivity is limited by the thermal losses in the coatings of the test masses. With improved coatings with lower loss, the thermal noise can be mitigated or cryogenic technology proposed in the Voyager detector can be implemented to improve both the broadband and the low frequency tuned sensitivity of the \SI{40}{\km} detector.}
  \label{fig:40km_lf}
\end{figure}

The most significant of these noises above ${\sim}\SI{20}{\Hz}$ is thermal noise in the test mass coatings. This noise is significant for the compact binary tuning up to ${\sim}\SI{50}{\Hz}$ for the \SI{40}{\km} detector and up to ${\sim}\SI{200}{\Hz}$ for the \SI{20}{\km} detector. It is especially important for the low frequency tuned configurations in which quantum noise is reduced below this thermal noise, as is shown in \cref{fig:40km_lf}. Reducing coating thermal noise is thus an important area of research for realizing the low frequency sensitivity. The Cosmic Explorer design assumes that the same optical coatings will be used as those used in the A+ upgrade to Advanced LIGO~\citep{Miller:2014kma} which targets a factor of two decrease in coating thermal noise. Promising candidates have been identified~\citep{Vajente2021}, though these coatings have not yet been realized. Crystalline AlGaAs coatings are a particularly promising option on the Cosmic Explorer time scale~\citep{Penn:19,Koch2019} which would allow Cosmic Explorer to surpass the low frequency sensitivity shown in \cref{fig:40km_lf}, though much research is needed to make them a reality.

The many low-frequency noises particularly important for the science discussed in \cref{sec:low_freq_tuning}---most significantly Newtonian gravity gradients, seismic, and thermal noise from the test mass suspensions---and the technological advances necessary to meet the Cosmic Explorer targets are discussed in detail in \citep{Hall:2020dps}.

An alternative technology using cryogenic silicon test masses and a \SI{2}{\um} laser~\citep{LIGO:2020xsf} has been identified as a potential upgrade to the baseline Cosmic Explorer technology of room-temperature fused silica test masses and a \SI{1}{\um} laser and could also be used should thermal effects in the baseline technology prove intractable~\citep{Evans:2021gyd}. There are several new considerations with this technology~\citep{Hall:2020dps}. First, the light traversing the substrates of the test masses experiences a phase noise due to the temperature dependence of the index of refraction which is a potentially significant low frequency noise source for the silicon technology, especially for the \SI{20}{\km} detector, due to silicon's larger thermorefractive coefficient and thermal conductivity. This thermorefractive noise is suppressed by a factor of $\sqrt{\mathcal{F}}$, however, which presents a trade off between low frequency sensitivity favoring large $\mathcal{F}$, and high frequency sensitivity favoring small $\mathcal{F}$ to minimize SEC loss. Second, the need to radiatively cool the cryogenic test masses imposes a strict heat budget~\citep{LIGO:2020xsf} which adds an additional constraint on how low $\mathcal{F}$ can be made to limit the power absorbed in the optics. This makes developing low absorption and high quality silicon substrates and optical coatings for \SI{2}{\um} light particularly important. Finally, manufacturing high quantum efficiency photodiodes for \SI{2}{\um} light is a critical area of R$\&$D necessary to minimize readout loss (c.f.\ \cref{fig:20km_pm_quantum}) for this technology.

\section{Discussions}
\label{sec:discussion}

\begin{deluxetable*}{p{35mm}p{115mm}}
\tablecaption{Summary of the advantages of Cosmic Explorer tunings targeting high and low frequencies discussed in this work.\label{tab:sumry}}
\tablehead{\colhead{Configurations} & \colhead{Key Results} }
\startdata
\vspace{1mm} High frequency tuning & 
    \begin{itemize}
        \item A \SI{20}{\km} post-merger optimized detector improves the chances of observing the post-merger signal, critical to the understanding of hot dense matter, see~\cref{subsec:pmr_tune}.
        \item The high frequency tuned CE (see \cref{subsec:pmr_tune} and \cref{subsec:tidal_tune}) improves the ringdown signal-to-noise for lighter black-holes and for the discovery potential of exotic objects~\citep{ Mazur9545,Chirenti_2007,PhysRevD.94.084016,PhysRevLett.116.171101,echoes}.
        \item A tidal optimized detector improves the measurement of tidal parameters of neutron stars at low redshifts, enabling an improved measurement of the cold equation of state using the loudest signals detected by Cosmic Explorer, see~\cref{subsec:tidal_tune}.
    \end{itemize} \\
\vspace{-4mm} Low frequency tuning & 
    \begin{itemize}
        \vspace{-5mm}
        \item Improves the detection prospects of heavier population of POP-III stars at high redshift, see~\cref{subsec:snr_imp}. 
        \item Improves the observational prospects of continuous waves sources below \SI{300}{\Hz}, see \cref{subsec:cw}.
        \item Improves the measurement of tidal parameters of neutron stars at high redshifts, enabling an improved measurement as a function of the age of the neutron stars, see \cref{subsec:tides_highz}.
        \item Improves the ringdown signal-to-noise for heavier black-holes, see \cref{subsec:grav_tune}.
    \end{itemize} \\
\enddata
\end{deluxetable*}

Tests of General Relativity, such as polarization measurements and precise tests of high-spin black-holes, require multiple detectors~\citep{PhysRevLett.116.221101,PhysRevD.100.104036,PhysRevD.103.122002}.
Moreover, three or more third-generation gravitational-wave detectors are required to localize the source in the sky and to measure the source distance precise enough to confirm the sources at high redshifts~\citep{Evans:2021gyd, Borhanian_2021}.
This suggests that two Cosmic Explorer facilities along with Einstein Telescope are necessary to achieve the key science goals of third-generation gravitational-wave detectors described here.

We assert that having at least one \SI{40}{\km} Cosmic Explorer detector is integral in achieving the key science goal of Cosmic Explorer as it outperforms a \SI{20}{\km} in all science goals other than the access to post-merger physics.
Research into mitigating SEC losses is key to the success of the \SI{20}{\km} Cosmic Explorer detector to achieve the science goals that depend on achieving improved high-frequency sensitivity. 
\citep{Borhanian_2021} and the \citep{Evans:2021gyd} assert that a network of third-generation detectors is indispensable. 
In particular, the precise determination of the source redshift and sky localization necessitates a network of three third-generation detectors---two Cosmic Explorer and Einstein Telescope. The key findings and benefits from tuning are summarized in the \cref{tab:sumry}.
Lastly, we note that we consider a handful of metrics to quantify the performance of different tuned configurations. We urge the broader gravitational wave astronomical community to perform in-depth analysis other than the SNR metric used in the study.

\section*{Acknowledgments}
The authors would like to thank Reed Essick and Daniel Brown for a careful review of the manuscript. 
VS and SB thank the National Science Foundation for support through award PHY-1836702 and PHY-1912536. 
DD is supported by the National Science Foundation as part of the LIGO Laboratory, which operates under cooperative agreement PHY-1764464. 
KK and ME thank the National Science Foundation for support through award PHY-1836814.
PL is supported by the Natural Sciences and Engineering Research Council of Canada (NSERC).
EDH is supported by the MathWorks, Inc. 
JR thanks the National Science Foundation for support through awards
PHY-1806962 and PHY-2110441.
BSS thanks the National Science Foundation for support through awards PHY-2012083, PHY-1836779 and AST-2006384. 

\newpage
\bibliographystyle{aasjournal}
\bibliography{PMR}

\end{document}